# Using Static and Dynamic Malware features to perform Malware Ascription

Jashanpreet Singh Sraw[a] and Keshav Kumar[b]

[a] Thapar Institute of Engineering and Technology, Patiala, Punjab, India
[b] Savitribai Phule Pune University, Pune, Maharashtra, India

Malware ascription is a relatively unexplored area, and it is rather difficult to attribute malware and detect authorship. In this paper, we employ various Static and Dynamic features of malicious executables to classify malware based on their family. We leverage Cuckoo Sandbox and machine learning to make progress in this research. Post analysis, classification is performed using various deep learning and machine learning algorithms. Using the features gathered from VirusTotal (static) and Cuckoo (dynamic) reports, we ran the vectorized data against Multinomial Naive Bayes, Support Vector Machine, and Bagging using Decision Trees as the base estimator. For each classifier, we tuned the hyper-parameters using exhaustive search methods. Our reports can be extremely useful in malware ascription.

## Introduction

In the near future, objects have to connect with each other which can result in gathering private sensitive data and cause various security threats and cybercrimes (1). Malware, an abbreviation of malicious software, comes in various types such as viruses, worms, trojans, ransomware, spyware, rootkits and more – all of which can target different platforms. Operating systems like Windows, Android, OS X, Linux and others are commonly known to be victims of malicious code. As of early 2019, weekly statistics from VirusTotal, a popular service that analyzes malicious content, indicates that approximately 33% of detected malware are Win32 EXEs (2). Windows alone, dominates almost 75% of the global desktop operating system market share (3). Because of the prevalence of PE malware, we have chosen to focus on analyzing data found in executables.

Antivirus vendors use a variety of naming schemes for file identification. While the naming format varies from vendor to vendor, most will include a family name. The family name, as described by Microsoft documentation (4), is a label used to group malware by common characteristics, including attribution to the same authors. For the purposes of our research, we will focus on attributing malicious executables to their corresponding malware families as a proxy for ground truth. In order to extract features from our samples, we take advantage of several malware analysis tools as described in the next section. These features are trained and tested on Naive Bayes and Support Vector Machine classifiers. In a follow up experiment, we convert our malware into grayscale and colored images to feed into a Convolutional Neural Network (CNN) for classification. In our experiment, we obtain results for only a subsample of these images and leave the rest for future experimentation.

## Previous Work

Some teams, (5), (6), (7), have conducted surveys detailing methods used in malware analysis. The reports detail many of the features useful for malware classification or attribution. Depending on the type of analysis performed, different features will be available.

<u>Static Analysis</u>

In static analysis, malware characteristics can be collected without execution of the code itself. For Portable Executable (PE) files, which are native to Windows, analysts often use tools that can present tabular views of PE header information, disassemble machine language, extract printable strings, de- termine file hashes, etc. Some features commonly investigated as part of static analysis include file hashes, strings, op code sequences, DLL imports, API calls, and other metadata found in the PE header (6). In malware visualization, researchers propose to view the contents of software in raw numerical form, such as binary, decimal, or hexadecimal, and convert these strings of numbers into images (8).

<u>Dynamic Analysis</u>

In contrast to static analysis, dynamic analysis involves execution of the malicious code. While this is much harder to scale than static analysis, monitoring how a binary interacts with an infected system can lead to more insights. Analysts are often interested in recording API and system calls, processes, modifications to system files and registries, network communication, etc. (5), (9).

<u>Hybrid Analysis</u>

Both static and dynamic analysis have their own limitations when conducted individually. For example, packers that com- press software can be used as an obfuscation tool to obscure contents of an executable. This will often necessitate the manual efforts of a malware analyst to conduct further static analysis on machine code. Dynamic analysis is not always successful since some types of malware require a certain duration to pass or a specific event to trigger its execution. Hybrid analysis, the combination of both static and dynamic analysis, can lead to more comprehensive views of malicious programs and enable researchers to gather a larger set of features for classification (6).

## Analysis for Malware Classification

The security landscape is constantly changing. As re- searchers develop new techniques to defend against hostile threats, adversaries are adapting their behavior to evade detection (10). In this paper, we are interested in features beneficial for classification of the latest malware threats. As such, we will analyze samples collected between the dates of Feb 2018 - Feb 2019 from a malware repository called VirusShare (11). The collection of features involves a hybrid approach where static features will come from reports generated by a free malware scanning service called VirusTotal (2). The binary contents of our PE files will further be converted into images, creating an additional feature space for attribution. Finally, a subsample of our malware will undergo dynamic analysis in an open-source malware analysis system called Cuckoo (12) where dynamic features will be gathered.

Virus Share

Our dataset of executables was collected from VirusShare, an online repository providing security researchers and professionals access to samples of malicious code (11). The folders containing the malware involved in this project were uploaded to the VirusShare repository beginning Feb. 2018 and ending in Feb. 2019. The contents of the folders were filtered for PE files, and all non-Windows malware were discarded. This left us with approximately 120,000 samples.

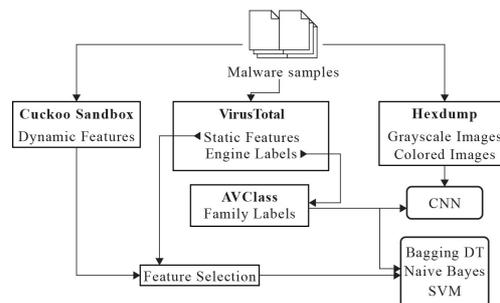

Figure 1. Hybrid analysis for malware classification

VirusTotal

Each of our 120,000 malicious samples were scanned through VirusTotal, generating a JSON report for each sample. From our collection of JSON reports, we gathered a set of static features for our experiments. In addition to providing static features, VirusTotal displays malware labels generated by each of its 69 AntiVirus scanners employed at the time of this writing.

AVCLass

The formatting and amount of information provided by a malware label varies from scanner to scanner. For example, Microsoft's malware labeling convention provides the type, targeted platform, family name, variant, and extra details in the form of a suffix (4). AVClass leverages the labels provided by the entire collection of antivirus scanners used by VirusTotal and outputs the most likely family name based on plurality voting. More details on AVClass can be found in the accompanying paper (13). When we test static features, we prune out families containing less than 20 samples. This leaves us with approximately 95,000 samples and 248 families.

Cuckoo Sandbox

We used stratified random sampling to select the executables that would undergo automated dynamic analysis in Cuckoo Sandbox. From our initial dataset of 120,000 samples belong- ing to 248 families, we consider only families with at least 50 samples, leaving us with 124 families. From each of these families, we randomly sample 20 malicious executables to submit for analysis by Cuckoo. This gives us 20 * 124 = 2,480 samples. Each sample is analyzed for two minutes and a behavioral report in JSON is generated. For the purposes of this experiment, we do not perform memory analysis on the system.

## Visualization

Applying a similar sampling strategy for visualization, we randomly sample 50 malicious executables from each of the 124 families containing at least 50 samples each. This gives us 50 * 124 = 6,200 malicious binaries we will consider for visualization experiments. We dump the raw bytes of each malware sample in unsigned decimal notation. We then map each decimal byte ranging from values 0-255 to a grayscale pixel with varying intensity. Black will be represented by the value 0 while white will be represented by 255. Each sample produces one image with width dependent on the executable's size. A reference table is provided below. This approach has been used in previous experiments such as (8) and [viv]. To fill the dimensions of the image with the appropriate number of pixels, some binaries required us to pad the end with byte values of "000". Some samples of our results and their corresponding family labels are shown in Figure 2.

Table 1. File size to image width for malware visualization

| File Size | Image Width |
|---|---|
| <10 kB | 32 |
| 10 kB - 30 kB | 64 |
| 30 kB - 60 kB | 128 |
| 60 kB - 100 kB | 256 |
| 100 kB - 200 kB | 384 |
| 200 kB - 500 kB | 512 |
| 500 kB - 1000 kB | 768 |
| >1000 kB | 1024 |

Separate colored images are also generated with the same raw bytes. Instead of mapping one byte value to one pixel, we propose to use a sliding window of three bytes (trigrams). The step sizes we use are one, in which case we will refer to the generated images as overlapping colored images, and three, in which case we refer to the generated images as non- overlapping-colored images. Each byte will represent the value of either a red, green, or blue pixel. For example, consider the following sequence of six bytes in base-10 or decimal notation:

077 090 144 000 003 000.

From this sequence, our overlapping window would generate four pixels of values: (R: 077, G: 090, B: 144), (R: 090, G: 144, B: 000), (R: 144, G: 000, B: 003), and (R: 000, G: 003, B: 000). Our non-overlapping window would simply generate two pixels of values: (R: 077, G: 090, B: 144) and (R: 000, G: 003, B: 000).

## Methodology

The features provided by each VirusTotal scan differs from sample to sample. The same can be observed from reports generated by the Cuckoo Sandbox. For this reason, we decided to gather feature counts for both VirusTotal reports and Cuckoo reports. We then identified the features occurring with the highest frequencies across our set of approximately 96,000 samples. Based on the feature descriptions and previous works, we deemed 13 recurring features as worth investigating.

## VirusTotal Features

- TRID: identification of file types from binary signature
- PE_RESOURCE_LIST: resource structure of a PE file
- EMBEDDED_DOMAINS_LIST: Domain names embedded in executable
- IMPORTS_LIST: DLLs imported and function calls used

- CONTACTED_URLS_LIST: external URLs contacted

Cuckoo Features

- SIGNATURE: extra behavioral context
- BEHAVIOR_CALLS: executables run and API calls used
- BEHAVIOR_DLL_LOADED: DLLs loaded
- NETWORK_HTTP: HTTP requests made
- NETWORK HOSTS: IP addresses contacted
- STRINGS: printable characters
- NETWORK_UDP_SRC: source IP/port of UDP communication
- NETWORK_UDP_DST: destination IP/port of UDP communication

Feature Extraction

The JSON reports were parsed, and features were extracted using the following methods:

- TRID: We only take the highest probable file type from TrID and use it as a feature.
- PE_RESOURCE_LIST: We combine each resource hash and its data type using a semi-colon (resource_hash:data_type)

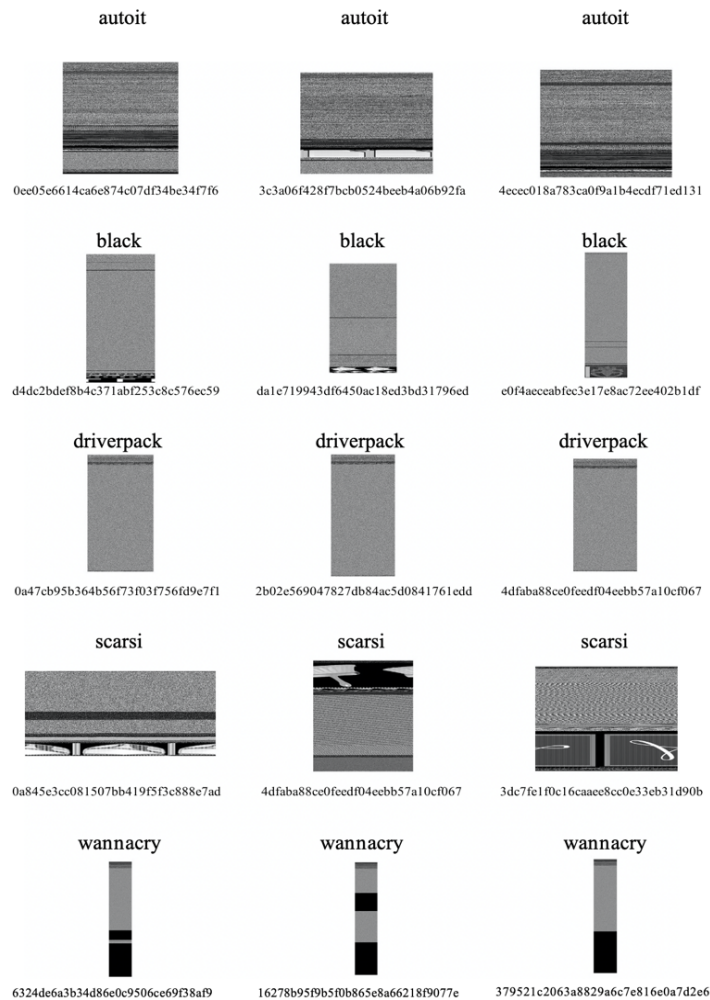

Figure 2. Grayscale malware images with family label

- EMBEDDED_DOMAINSLIST: We extract each embedded URL. Only the base URLs were extracted, and instances of [http://, https://, www.] were stripped from the URL.
- IMPORTS_LIST: We combine each DLL with each of its function calls using a semi-colon (import_name:method).
- CONTACTED_URLS_LIST: We extract each contacted URL. Only the base URLs were extracted, and [http://, https://, www.] were stripped from the URL.
- SIGNATURES: We extract each signature from the report.
- BEHAVIOR_CALLS: We extract each API call that an executable run in the session.
- BEHAVIOR_DLL_LOADED: We extract each DLL that an executable loads.
- NETWORK_HTTP: We extract each HTTP request.
- NETWORK_HOSTS: We extract each HOST request.
- STRINGS: We extract each string. All spaces were removed from each string.
- NETWORK_UDP_SRC: We combine each IP with its port using a semi-colon (IP:port).
- NETWORK_UDP_DST: We combine each IP with its port using a semi-colon (IP:port).

Not all features will exist for every malware report. During feature extraction, if the feature does not exist, then in most cases we will simply skip that feature. In some scenarios we will skip that particular malware sample entirely.

Vectorization

The types of vectorizations used by each feature were deter- mined from preliminary testing. We chose the types that were most suited to the feature and provided the highest accuracy possible. The following are the vectorization used for each feature:

- TRID: One-hot-encoded
- PE RESOURCE_LIST: 1, 2, 3 Grams
- EMBEDDED_DOMAINS_LIST: 1 Gram
- IMPORTS_LIST: 1, 2, 3 Grams
- CONTACTED_URLS_LIST: 1, 2, 3 Grams
- SIGNATURES: 1, 2, 3 Grams
- BEHAVIOR_CALLS: 3 Grams
- BEHAVIOR_DLL_LOADED: 1, 2, 3 Grams
- NETWORK_HTTP: 1, 2, 3 Grams
- NETWORK_HOSTS: 1, 2, 3 Grams
- STRINGS: 1, 2, 3 Grams.
- NETWORK_UDP_SRC: 3 Gram
- NETWORK_UDP_DST: 3 Gram

After all features were extracted, we employed a Chi-squared test to rank each feature. We then picked the top 50% of ranked features as our features for classification.

Feature Contribution Experiments

We use two sets of features for two different experiments. In the first experiment, we consider VirusTotal features only. In the second experiment, we consider the combination of both VirusTotal and Cuckoo features. Using methods similar to those employed in (15), we conduct empirical testing on each static and dynamic feature to understand how they contribute to classification results. For each of the features, we select one to leave out while using the rest as input for the following classifiers: (1) Multinomial Naive Bayes, (2) Support Vector Machine, and (3) Bagging using Decision Trees as the base estimator. We measure Accuracy, Recall, Precision, F-score, and Execution Time. The results of our experiments are shown under section V, subsections C and D. Preliminary tests with the features NETWORK UDP SRC and NETWORK UDP DST proved they were non-useful and highly detrimental. Therefore, they were dropped from any further testing. Execution time was shown to be too volatile between each leave-out run due to the nature of processing. Therefore, we do not consider Time when evaluating feature contributions.

Classifiers

Using the features gathered from VirusTotal (static) and Cuckoo (dynamic) reports, we ran the vectorized data against three different classifiers: (1) Multinomial Naive Bayes, (2) Support Vector Machine, and (3) Bagging using Decision Trees as the base estimator. For each classifier, we tuned the hyper-parameters using exhaustive search methods.

For SVMs, we used the linear kernel, which produced the most consistent and best results. Other kernels we attempted to use were: rbf, poly, and sigmoid. For Bagging, the hyper-parameters for the best results with a reasonable execution time were found to be: n estimators=1000, max features=0.1, and max samples=0.1. We found that SVM produced the best results over Naive Bayes and Bagging. Therefore, we only show our final results using SVM.

Visualization

For our visualization experiments, it was necessary to resize our images with normalized dimensions before passing them into our CNN. Our sample size was n = 6, 200 sampled binaries. Let $m_k$ with 1 k 6, 200 denote a set of images corresponding to a single malware sample. Then our entire collection of image sets is represented by:

$$M = \{m_1, m_2, ..., m_n\}$$

For each malware image set m⊂ M, we generated nine corresponding normalized images. Then,

$$m_k = \{g_c, g_m, g_e, o_c, o_m, o_e, n_c, n_m, n_e\}$$

where:
- $g_c$ = grayscale, compressed image
- $g_m$ = grayscale, median image
- $g_e$ = grayscale, expanded image
- $o_c$ = overlapping coloured, compressed image
- $o_m$ = overlapping coloured, median image
- $o_e$ = overlapping coloured, expanded image
- $n_c$ = non-overlapping coloured, compressed image
- $n_m$ = non-overlapping coloured, median image
- $n_e$ = non-overlapping coloured, expanded image

To derive compressed image dimensions, we took the minimum of the square root of pixel dimensions from one of the sets of grayscales, non-overlapping colored, or overlapping colored images. We then applied the floor function on the result. This gives us the dimensions of the two sides of the normalized compressed image. Let $I_g$ denote the set of grayscale image dimensions and i be an arbitrary pair of grayscale image dimensions within the set. Then the compressed height, $h_c$, and compressed width, $w_c$, for all images within the set $I_g$, are derived as follows:

$$h_c = w_c = \min_{i \in I_g}(\sqrt{i}) \quad (1)$$

The same method is applied for non-overlapping colored and overlapping colored images.

To derive median image dimensions, we took the square root of the median dimension from one of the sets of grayscales, non-overlapping colored, or overlapping colored images and applied the floor function on the result. This gives us the dimensions of the two sides of the normalized median image. The median height, $h_m$, and median width, $w_m$, for all images within the set $I_g$, are derived as follows:

$$h_m = w_m = \sqrt{m} \quad (2)$$

The same method is applied for non-overlapping colored and overlapping colored images. To derive compressed image dimensions, we took the maximum of the square root of pixel dimensions from one of the sets of grayscales, non-overlapping colored, or overlapping colored images. We then applied the ceiling function on the result. This gives us the dimensions of the two sides of the normalized expanded image. The expanded height, $h_e$, and expanded width, $w_e$, for all images within the set $I_g$, are derived as follows:

$$h_c = w_c = \max_{i \in I_g}(\sqrt{i}) \quad (3)$$

The same method is applied for non-overlapping colored and overlapping colored images. The results of normalizing our sets of image dimensions are shown in Table II. Using Keras (16) and the dimensions derived in the previous sections, we used nearest neighbor interpolation to resize our original images in memory. Our image generator randomly samples, shuffles, and augments images to expand the initial training set. These images are then fed into a 2D Convolutional Neural Network with a multi-class output where output nodes represent the probability of a sample belonging to 1 of 124 malware families. We use categorial cross entropy for our loss function and recorded results for accuracy. Results are listed in section V under subsection E.

**Results**

Machine Specifications

The following are the machine specifications used to produce the results in Tables III, IV, V, VI, VII, and VIII:
- Intel(R) Core(TM) i7-7700K CPU @ 4.20 GHz
- 16 GB RAM

## Metrics

The metrics which we use to evaluate our classifiers are defined as follows:

$$\text{Accuracy} = \frac{TP + TN}{TP+TN+FP+FN} \quad (4)$$

$$\text{Precision} = \frac{TP}{TP+FP} \quad (5)$$

$$\text{Recall} = \frac{TP}{TP+FN} \quad (6)$$

$$\text{F1-Score} = 2 * \frac{Precision*Recall}{Precision+Recall} \quad (7)$$

where TP = True Positives, FP = False Positives, and FN = False Negatives.

Table 2. Normalized image dimensions

| Image Set | Compressed (px) | Median (px) | Expanded (px) |
|---|---|---|---|
| Grayscale | 56 x 56 | 909 x 909 | 5,774 x 5,774 |
| Nonoverlapping RGB | 32 x 32 | 525 x 525 | 3,334 x 3,334 |
| Overlapping RGB | 52 x 52 | 850 x 850 | 5,402 x 5,402 |

## VirusTotal Feature Contributions

Our investigation of static features involved around 95,000 samples from a total of 248 malware families. The results of using all VirusTotal features for the SVM are shown in Table III. We chose one static feature to leave out and re-ran our SVM with the rest of the features as input, repeating this experiment for each of the five features. The results are shown in Table VII. None of the static features were determined to be non-useful and therefore all five features were used for the final classifier. As such, our final results remained the same, with the exception of Time which had a negligible change. Using all static features, the final accuracy was shown to be about 85%.

Table 3. VirtualTotal using SVM (95,000 samples)

| Accuracy | Precision | Recall | F-score | Time (s) |
|---|---|---|---|---|
| 84.99 | 83.98 | 84.99 | 83.72 | 3341 |

## Cuckoo Feature Contributions

For our investigation of dynamic features, not all malware samples contained all of the eight features of interest. From our batch of 2,480 Cuckoo reports, we chose to include reports from all malware families containing at least 18 samples. This leaves us with 1,936 samples from 101 malware families. Our initial results, considering only static VirusTotal features in our SVM are shown in Table IV. The results of combining all of the Cuckoo features with the VirusTotal features for the SVM are shown in Table V. The difference between the two tables shows how hybrid analysis (combining both static and dynamic features), can improve accuracy in attribution of malware to their families. Like our previous experiments on the 95,000 sample dataset, we chose one dynamic feature to leave out and re-ran our SVM with the rest of the features as input. We repeated this experiment for each of the eight dynamic features. The results of this experiment are shown in Table 8. NETWORK HTTP and NETWORK HOSTS were determined to be non-useful. By dropping these two features, our final accuracy increased by 0.11% to 67.87%. The final results are displayed in Table 6.

**Table 4.** VirusTotal features using SVM: initial results (1,936 samples)

| Accuracy | Precision | Recall | F-score | Time (s) |
|---|---|---|---|---|
| 61.73 | 64.57 | 61.73 | 60.69 | 41 |

**Table 4.** VirusTotal and cuckoo features using SVM: initial results (1,936 samples)

| Accuracy | Precision | Recall | F-score | Time (s) |
|---|---|---|---|---|
| 67.87 | 69.72 | 67.87 | 66.48 | 1462 |

**Table 5.** VirusTotal and cuckoo features using SVM: final results (1,936 samples)

| Accuracy | Precision | Recall | F-score | Time (s) |
|---|---|---|---|---|
| 67.98 | 69.79 | 67.98 | 66.66 | 1946 |

Classifying Malware Images

For each of the 6,200 malware samples involved in the visualization experiment, we generated nine associated normalized images – three of which were grayscale and six of which were colored. Due to time constraints, our classification results with the 2D CNN are limited to compressed grayscale images. For compressed grayscale images, accuracy was 97%, proving that features from malware visualization can be used for attribution.

**Conclusion**

In this paper, we have conducted a hybrid analysis of malware features generated by VirusTotal reports, Cuckoo Sandbox analysis, and binary visualization. Our performance metrics evaluated how well these features trained our classifiers to attribute malicious binaries to malware families. Using a Support Vector Machine with features generated by Virus- Total and Cuckoo gave the best results, proving to have higher performance when compared to Naive Bayes and Bagging using Decision Trees. In addition, experiments were conducted to evaluate the efficacy of each static and dynamic feature for attribution. All static features were shown to useful and only a few dynamic features had detrimental contributions to our performance metrics. Finally, we performed malware image classification using a 2D Convolutional Neural Network. Our model was able to classify malware based on compressed grayscale images with high accuracy, confirming that features within malware images can be used for attribution.

**Table 6.** VirusTotal features using SVM: leave-one-out results

| Dropped Feature | Accuracy | Precision | Recall | F-score | Time (s) |
|---|---|---|---|---|---|
| TRID | 84.81 | 83.76 | 84.81 | 83.48 | 3028 |
| PE RESOURCE LIST | 83.99 | 83.2 | 83.77 | 82.45 | 2117 |
| EMBEDDED DOMAINS LIST | 84.73 | 83.71 | 83.43 | 83.43 | 3527 |
| IMPORTS LIST | 80.64 | 79.65 | 80.64 | 78.69 | Z618 |
| CONTACTED URLS LIST | 84.78 | 83.81 | 84.78 | 83.37 | 3271 |

**Table 7.** VirusTotal and cuckoo features using SVM: leave-one-out results

| Dropped Feature | Accuracy | Precision | Recall | F-score | Time (s) |
|---|---|---|---|---|---|
| TRID | 67.87 | 69.92 | 67.87 | 66.49 | 1231 |
| PE RESOURCE LIST | 67.20 | 69.27 | 67.2 | 66.00 | 1290 |
| EMBEDDED DOMAINS LIST | 67.39 | 69.62 | 67.36 | 66.10 | 1346 |
| IMPORTS LIST | 66.99 | 69.27 | 66.99 | 65.78 | 1209 |
| CONTACTED URLS LIST | 67.72 | 69.4 | 67.72 | 66.33 | 1528 |
| SIGNATURES | 66.12 | 68.19 | 66.12 | 64.71 | 1525 |
| BEHAVIOR CALLS | 67.46 | 69.58 | 67.46 | 66.19 | 1137 |
| BEHAVIOR DLL LOADED | 67.56 | 69.38 | 67.56 | 66.19 | 1333 |
| NETWORK HTTP | 67.92 | 69.60 | 67.92 | 66.47 | 1432 |
| NETWORK HOSTS | 67.92 | 69.60 | 67.92 | 66.47 | 1217 |
| STRINGS | 67.39 | 69.37 | 67.36 | 66.17 | 112 |

## Future Work

Though there has been work done in the past in clustering with various features of malware such as those found in Cuckoo reports, [17], [18], we believe there is room for future work in unsupervised clustering for malware images. Many papers exploring malware visualization instead focus on the task of classification. However, with an unlabeled dataset, features can be extracted from a CNN and be used as input for various clustering algorithms. Analyzing similarities among clustered malware may provide useful benefits such as signature generation or a better understanding of malware lineage [18].